\documentclass[prb,aps,a4paper]{revtex4}

\usepackage{graphicx}

\newif\ifpdf
\ifx\pdfoutput\undefined
\pdffalse 		
\else
\pdfoutput=1 	
\pdftrue
\fi

\begin{document}

\title{Graphene buffer layer on Si-terminated SiC : a multi-minima energy surface studied
with an empirical interatomic potential}

\author{Evelyne Lampin}
\email[]{evelyne.lampin@isen.iemn.univ-lille1.fr}
\author{Catherine Priester}
\author{Christophe Krzeminski}
\affiliation{IEMN - BP 60069 - 59652 Villeneuve d'Ascq Cedex - France}
\author{Laurence Magaud}
\affiliation{Institut N\'eel - BP 166 - 38042 Grenoble Cedex 9 - France}

\date{\today}

\begin{abstract}
The atomistic structure of the graphene buffer layer on Si-terminated SiC is studied using a modified environment-dependent interatomic potential (EDIP). The investigation of equilibrium state by conjuguate gradients suffers from a complex multi-minima energy surface. A dedicated procedure is therefore presented to provide a suitable initial configuration on the way to the minimum. The result forms an hexagonal pattern with unsticked rods to release the misfit with the surface. The structure presents an agreement with the global pattern obtained by experiments and even with the details of an \emph{ab initio} calculation. 
\end{abstract}


\maketitle

\section{Introduction}
Graphene is an attractive alternative to silicon for high-mobility devices due to its unique electronic properties. The graphitisation of SiC by high-temperature annealing \cite{Berger04} is often seen as the most promising route to fabricate large layers of graphene. Although many experimental works were devoted to the refinement of the fabrication process, the graphitization mechanism remains poorly understood. In particular, the control of the sublimation of Si atoms during and after formation of the first graphene layer is a key step of the fabrication of larger and more regular layers \cite{Sutter09}. \emph{ab initio} calculations have shown that the first graphene layer, the so-called buffer layer, on the Si-terminated surface has not a graphenelike dispersion \cite{Varchon07} and is strongly bonded to the substrate \cite{Varchon08},\cite{Kim08}. The experimental counterpart is not simple, in particular due to the high sensitivity to the preparation procedure \cite{Riedl07}. Further understanding could be provided with molecular dynamics calculations that describe the evolution of an atomic system submitted to thermal annealing. Such calculations could only be carried out with rough approximations in the framework of \emph{ab initio} descriptions and the use of an interatomic potential is therefore more adapted. We present in this study the results of calculations performed with a version of the environment-dependent interatomic potential (EDIP) modified for SiC by Lucas \cite{Lucas06}. The equilibrium state of the buffer layer on SiC(0001) is studied. The difficulty to achieve convergence in this multi-minima energy surface is demonstrated, and a specific procedure is given to reach equilibrium. The configuration of minimal energy presents the symetries observed by experiments. Moreover, the details are compared to \emph{ab initio} results. A good agreement is obtained and the optimum even presents an higher level of order. We therefore conclude on the ability of the potential to describe the graphitisation of SiC surface. 

\section{Method}
\begin{figure}[htb]
\begin{minipage}[c]{0.45\linewidth}
\includegraphics[width=0.95\linewidth]{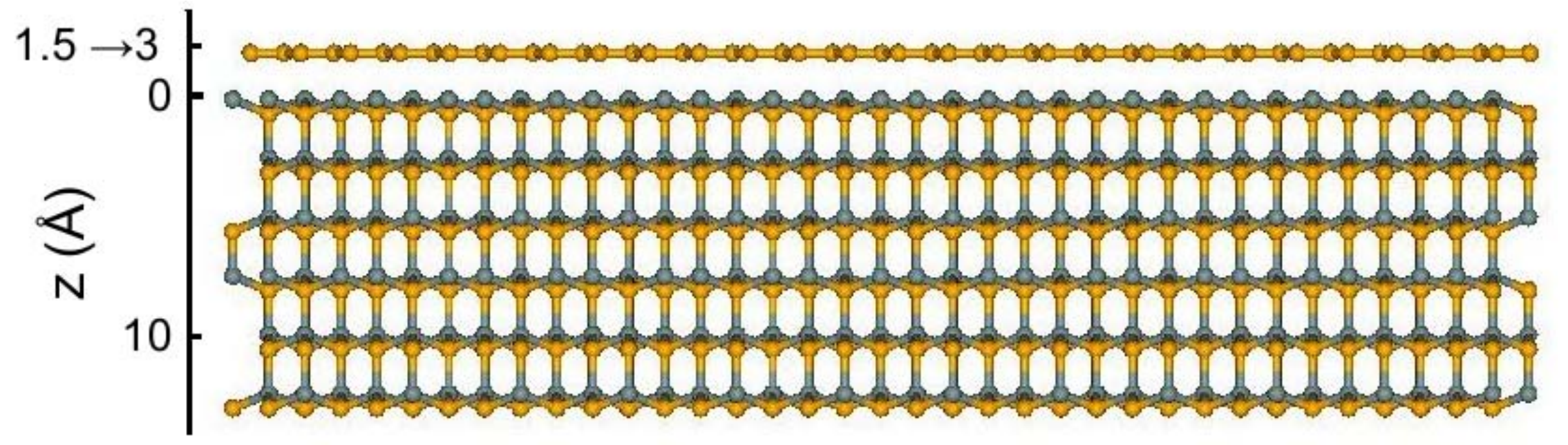}
\end{minipage}
\begin{minipage}[c]{0.45\linewidth}
\includegraphics[width=0.95\linewidth]{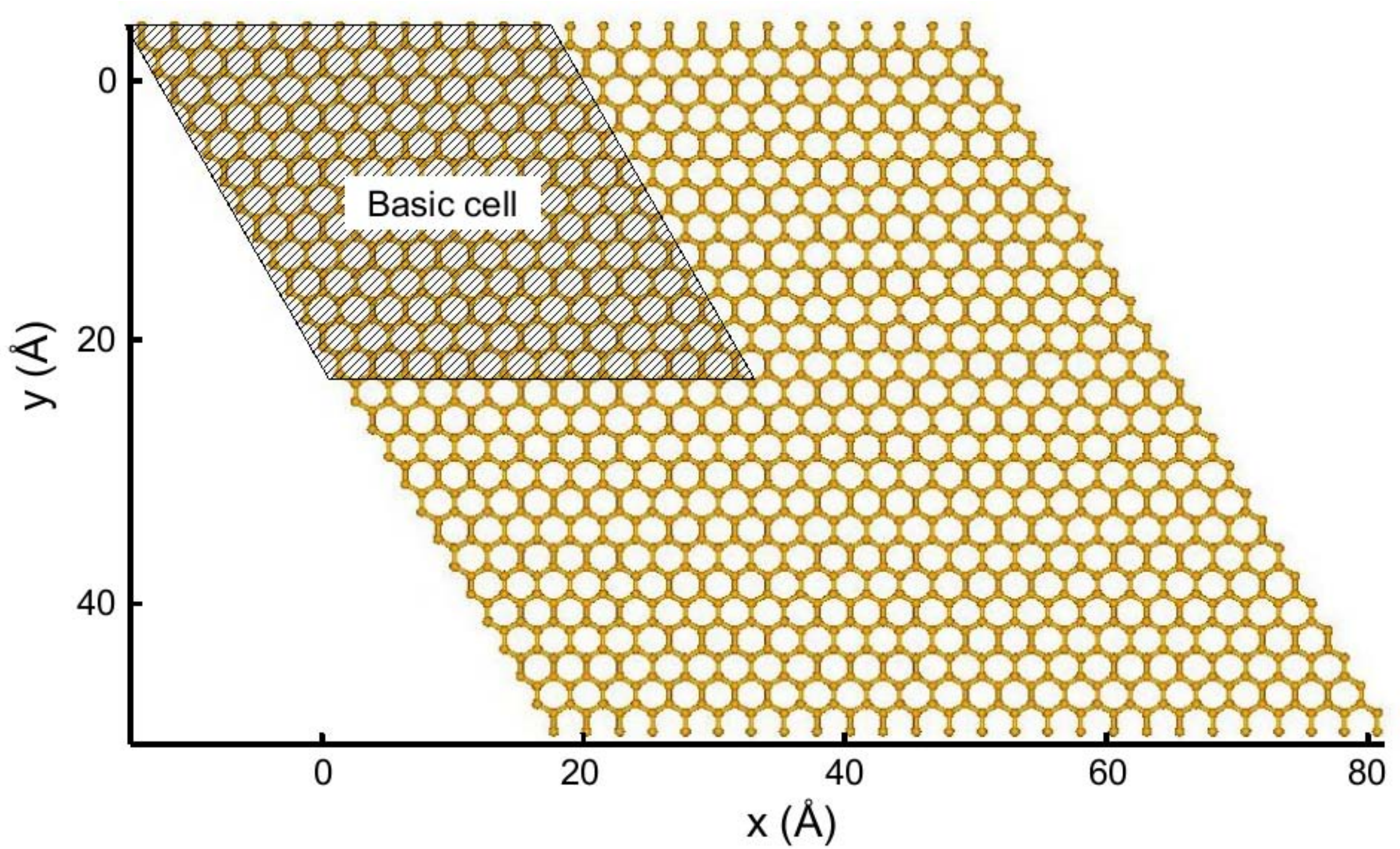}
\end{minipage}
\caption{\label{FigGraphonSiC} Lateral (left) and top (right) views of the atomic system under study. The C atoms are colored in orange, the Si in grey. The delimitations of the basic cell are shown on the right view.}
\end{figure}

In a first step, a model of graphene on SiC(0001) is constructed. For the sake of simplicity we consider a 3C-SiC substrate, but the system is perfectly equivalent to 4H- and 6H-SiC provided a planar SiC surface is considered (the equivalence does not hold in the case of a step for example). The lattices of graphene and SiC do not match. A low misfit is obtained for a $(6\sqrt{3}\times6\sqrt{3})R30^{\circ}$ supercell \cite{Forbeaux98} put in regard of a sheet of 13-hexagon wide honeycomb. The misfit, defined by the ratio~:
\begin{equation}
\frac{13 \times a_\mathrm{graphene}}{6\sqrt{3} \times a_\mathrm{SiC}}
\end{equation}
is equal to 0.7 \% experimentally and 3.0 \% with the lattice parameters ($a_\mathrm{graphene}$, $a_\mathrm{SiC}$) obtained at equilibrium using the interatomic potential presented hereafter. The graphene layer of the basic cell contains 338 atoms of C. The depth of the SiC substrate has to be high enough to absorb the strain generated by misfit. We choose to use 6 SiC bilayers, with a total of 1296 atoms, although the dependence of the results on the depth of the substrate will be tested in the last section of this article. The basic hexagonal cell is replicated twice along $x$ and $y$ to facilitate the visualization of the patterns formed by atoms in the buffer layer. Fig. \ref{FigGraphonSiC} presents lateral and top views of the corresponding sytems of 6536 atoms.

The interactions between atoms are described by an empirical interatomic potential. Empirical interatomic potentials are developed for a specific domain of a material physics. It is generally difficult to estimate the range of validity of a given potential when it is applied to a different phenomenon for the first time \cite{Krzeminski07}. Indeed one of the aim of this work is to confront the potential under study to \emph{ab initio} calculations in order to validate its application to the study of graphene on SiC. The interatomic potential is a slightly modified version of EDIP \cite{Bazant96}, \cite{Bazant97}, \cite{Justo98}  parametrised for C-C and Si-C interactions by Lucas \cite{Lucas06}. Other classical interatomic potentials such as Tersoff \cite{Tersoff88} and Brenner \cite{Brenner90} have also been adapted and parametrised for Si-C \cite{Tersoff89}, \cite{Dyson96}, \cite{Devanathan98}, \cite{Gao02}, \cite{Erhart05}. Tewary and Yang \cite{Tewary09} have constructed a parametric interatomic potential but it is restricted to graphene (only C-C interations). As for the study of SiC graphitization, Tang et al. used Tersoff's potential in molecular dynamics simulation of the heating of Si-less layers on SiC \cite{Tang08_1} and of graphene nanoribbons \cite{Tang08_2} although the coordination of order 3 is not natural for a potential developed for diamond lattice. In the present study,  the environment-dependent mathematical formulation of EDIP 
 has been assumed to be an advantage for the modelisation of the bonding of graphene layer on the SiC surface including $sp2$ and $sp3$ bonds. The potential energy of a system of atoms at position $\vec{r}_i$ is given by~:
\begin{eqnarray}
E = \sum_{i} E_i,\nonumber\\
 E_i=\sum_{j\ne i}V_2(\vec{r}_{i},\vec{r}_j,Z_i) + \sum_{j\ne i}\sum_{k\ne i,k>i}V_3(\vec{r}_{i},\vec{r}_{j},\vec{r}_{k},Z_i).
\end{eqnarray}
The coordination number is~:
\begin{eqnarray}
Z_i = \sum_{m\ne i}f(r_{im}), f(r)=\left\{
\begin{array}{cc}
1, &r<c\\
\exp\left[\alpha/\left(1-\left(\frac{a-c}{r-c}\right)^3\right)\right]\Delta(r), &c<r<a\\
0,&r>a.
\end{array}
\right.
\end{eqnarray}
The corrective function $\Delta(r)$ is added by Lucas \cite{Lucas06} to avoid the second nearest neighbour Si-Si interactions in SiC to be accounted for. The function is equal to unity except if two Si atoms are interacting~:
\begin{equation}
\Delta(r)=\left\{
\begin{array}{cc}
1, &r<c\\
\exp\left[\alpha/\left(1-\left(\frac{a-c}{r-c}\right)^3\right)\right], &c<r<a-\delta\\
0,&r>a-\delta
\end{array}
\right.
\end{equation}
with $\delta=0.3$ \AA. The two-body part is given by ~:
\begin{equation}
V_2(\vec{r}_{i},\vec{r}_j,Z_i) = A \left[\left(\frac{B}{r_{ij}}\right)^{\rho} - \exp(-\beta Z_i^2)\right] \exp\left(\frac{\sigma}{r_{ij}-a}\right)\Delta(r_{ij}).
\end{equation}
The three-body term is made of radial and angular contributions~:
\begin{eqnarray}
V_3(\vec{r}_{i},\vec{r}_{j},\vec{r}_{k},Z_i) = \exp\left(\frac{\gamma}{r_{ij}-a}\right)\Delta(r_{ij})\exp\left(\frac{\gamma}{r_{ik}-a}\right)\Delta(r_{ik})h(\cos\theta_{ijk},Z_i),\nonumber\\
h(l,Z)=\lambda\left[(1-\exp[-Q(Z)(l+\tau(Z))^2])+\eta Q(Z)(l+\tau(Z))^2\right],\\
Q(Z)=Q_0\exp(-\mu Z), \tau(Z)=u_1+u_2(u_3e^{-u_4Z}-e^{-2u_4Z}),\nonumber
\end{eqnarray}
with $u_1=-0.165799$, $u_2=32.557$, $u_3=0.286198$ and $u_4=0.66$. The parameter set is given in Tables \ref{TabParam2} and \ref{TabParam3}. 
\begin{table}[htb]
\begin{tabular}{cccc}
\hline\hline
& Si-Si & C-C & Si-C\\
\hline
a(\AA)&3.1213820 & 2.2918049& 2.6746212\\
c(\AA)& 2.5609104& 1.5698566& 2.0050591\\
$\alpha$&3.1083847 & 1.1268088& 2.1175967\\
A(eV)& 7.9821730& 11.6773355&9.6545591 \\
B(\AA)& 1.5075463& 0.9612016& 1.2037674\\
$\rho$& 1.2085196& 2.8779528& 2.0432362\\
$\beta$& 0.0070975& 0.0255960& 0.0163467\\
$\sigma$(\AA)& 0.5774108& 0.7103453& 0.6404382\\
$\gamma$(\AA)& 1.1247945& 1.2258251& 1.1742237\\
\hline\hline
\end{tabular}
\caption{\label{TabParam2}Two-body parameters of the EDIP potential used in the present work \cite{Justo98},\cite{Lucas06}.}
\end{table}
\begin{table}[htb]
\begin{tabular}{ccccccc}
\hline\hline
& Si-Si-Si & C-C-C & Si-SiC&C-SiC& Si-CC\\
& & & or -CSi&or -CSi& or C-SiSi\\
\hline
$\lambda$(eV)& 1.4533108& 1.7800184&1.5349877 & 1.6983415& 1.6166646\\
$\eta$& 0.2523244& 0.8115872& 0.3921401&0.6717715&0.5319558\\
Q$_0$&312.1341346 & 417.5476826&338.4875216 & 391.1942956&364.8409086\\
$\mu$ &0.6966326& 0.6122148&0.6755281 &0.6333192 &0.6544237\\ \hline\hline
\end{tabular}
\caption{\label{TabParam3}Three-body parameters of the EDIP potential used in the present work \cite{Justo98},\cite{Lucas06}.}
\end{table}The result is a potential able to describe Si- and C-based crystals and their defects. The only noticeable drawback, as in standard \emph{ab initio} calculations, is that the Van der Waals interactions are poorly described. Therefore graphite could only be obtained by fixing the distance between two planes of graphene.

Periodic boundary conditions are applied in the lateral directions $x$ and $y$ (see Fig. \ref{FigGraphonSiC}. The bottom SiC bilayer in the $z$ direction is frozen to avoid reconstruction of the C-terminated surface. The relaxation is performed with the method of the conjugate gradients \cite{Payne92} until a precision of 10$^{-8}$ is reached on the total energy (tenths of $\mu$eV/atom) after tenths to hundreds of steps. The initial gap between graphene and the surface is set between 1.5 and 3 \AA\ by step of 0.01 \AA, the cutoff of the Si-C interactions being equal to 2.67 \AA\ and the equilibrium Si-C distance to 1.91 \AA. Three lateral locations of the graphene plane compared to the SiC surface are tested :
\begin{itemize}
\item shift 1 : a C atom in graphene at the vertical of a Si surface atom
\item shift 2 : center of a graphene hexagon at the vertical of a Si surface atom
\item shift 3 : middle of a C-C bond in graphene at the vertical of a Si surface atom.
\end{itemize}

\section{Results}
\begin{figure}[htb]
\includegraphics[width=0.5\linewidth]{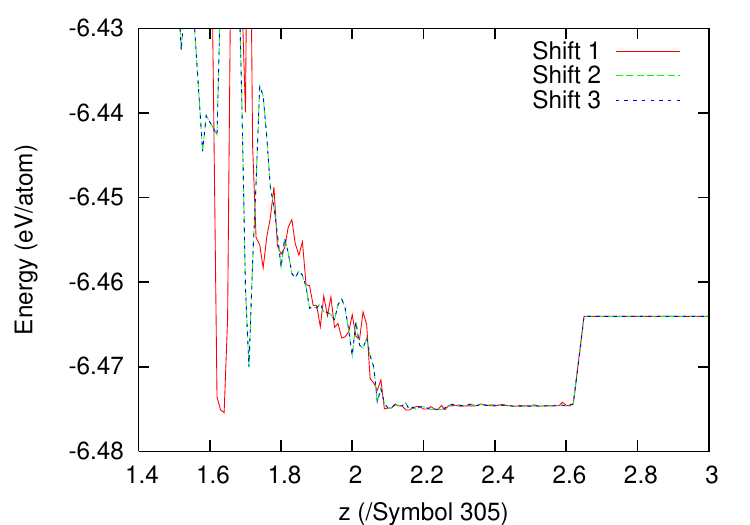}
\caption{\label{FigCherche} Total potential energy versus initial gap between graphene and surface.}
\end{figure}

The total energy after relaxation is presented in Fig. \ref{FigCherche} as a function of the gap between the plane of graphene and the surface and for the three shifts.
The second and third shifts, corresponding to an alignment on the center of an hexagon or on the middle of a bond, give exactly the same energy. Indeed these configurations should be equivalent. In the three cases, the result strongly depends on the initial gap. Above the Si-C cutoff radius of the interactions (2.67 \AA), the energy does not depend on the initial gap as expected and corresponds to the sum of the independent contributions of the SiC substrate and of the plane of graphene. In the range $[$2.1-2.6$]$\AA, the energy only slighlty evolves with the initial gap while there is a strong variation of the result when the graphene plane is set at less than 2.1~\AA\ of the SiC surface. From one value of the gap to the neighboring one, the bonds between graphene and Si surface involve different atoms and this is enough to break bonds and form new ones. This is the explanation of the great dependence of the total energy on the initial gap of graphene, and the presence of multiminima in energy. The method of conjuguate gradient is obviously not able to allow the system to go out of a local minimum. The open question is therefore whether the absolute minima is found or not with this method.

\begin{figure}[htb]
\begin{minipage}{0.45\linewidth}
\includegraphics[width=0.95\linewidth]{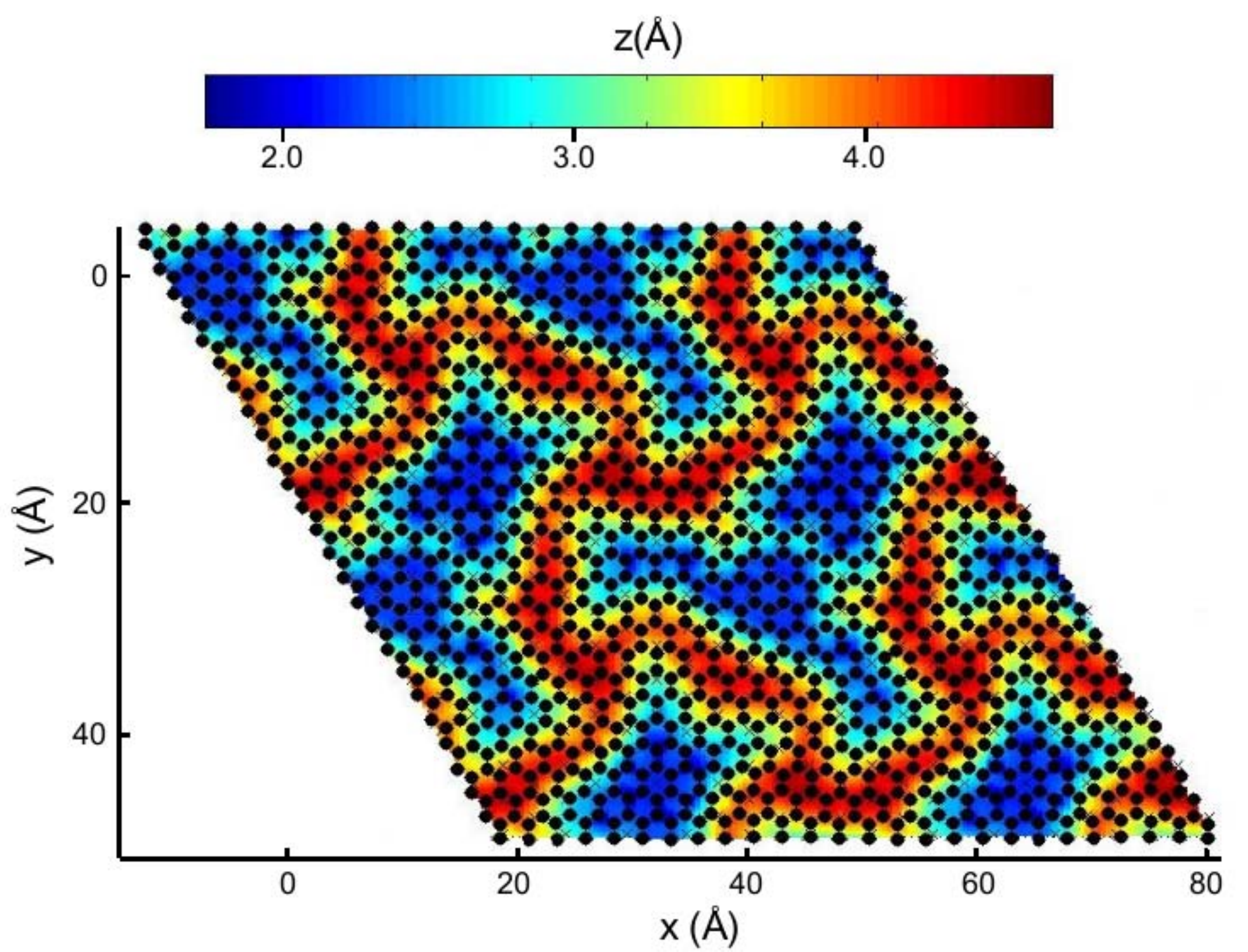}
\caption{\label{Fig1.63} Color map of the position of the C atoms in graphene along $z$ (black dots). Initial gap set to 1.63 \AA. Blue (red) means near (far from) the SiC surface. Light crosses indicate the position of the Si atoms of the surface.}
\end{minipage}
\begin{minipage}{0.45\linewidth}
\includegraphics[width=0.95\linewidth]{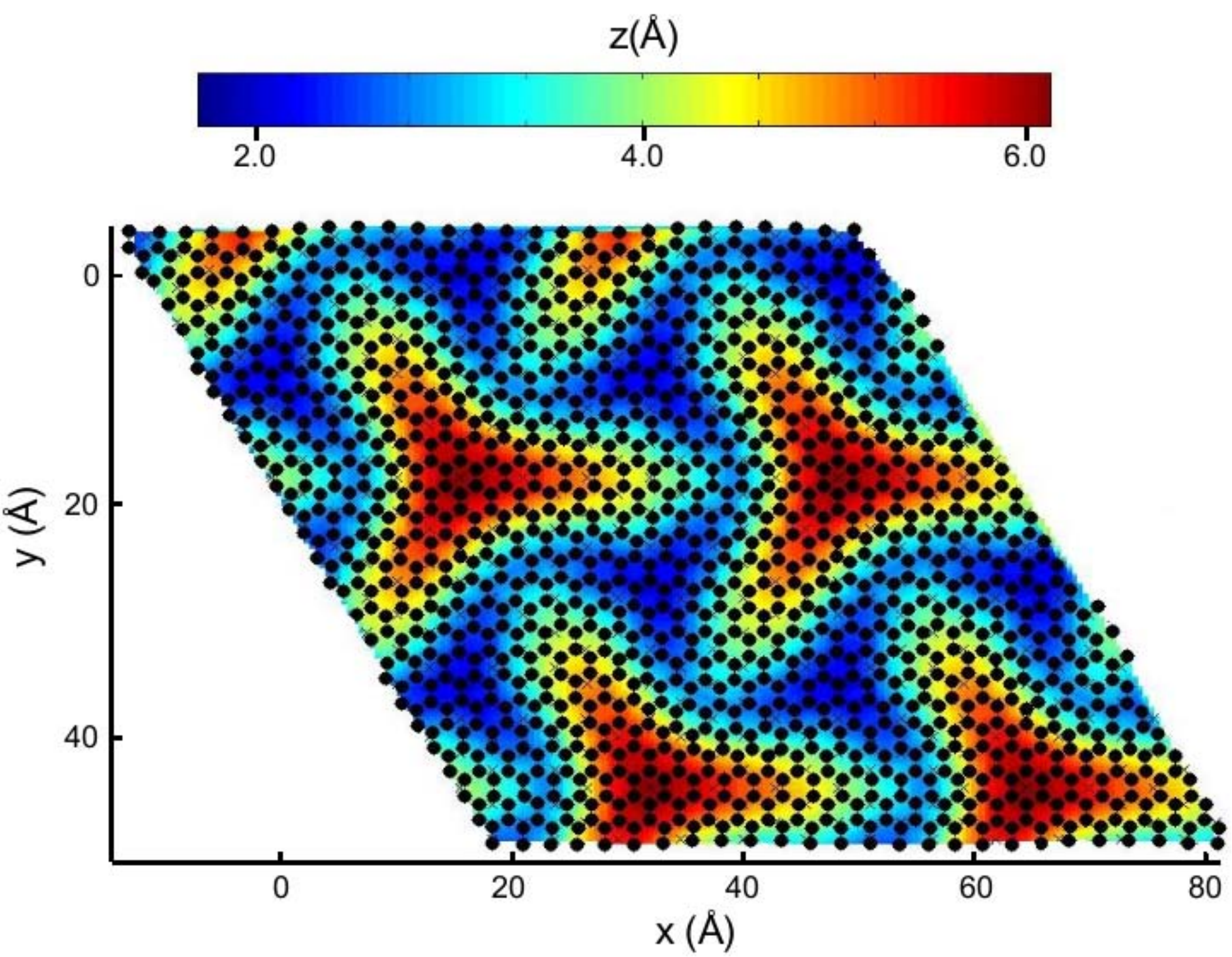}
\caption{\label{Fig2.26} Same figure than Fig. \ref{Fig1.63} but for an initial gap of 2.26 \AA.}
\end{minipage}
\end{figure}

Moreover, we studied in more details the structure of the graphene plane and its bonding to the surface for the two gaps that give a low energy in the configuration of shift 1, i. e. an alignment of a C in graphene on a Si at the surface. These gaps are 1.63 and 2.26 \AA. Color maps of the position of the C atoms along $z$ superimposed on dots that figure the C  position are shown on Figs. \ref{Fig1.63} and \ref{Fig2.26} respectively.
The patterns have an order 3 symmetry, although an order 6 could be expected, and the structure is surprisingly not regular in Fig. \ref{Fig1.63}. Clearly, the minimisation does not seem achieved. The two figures however evidence the complexity of the energy surface, since the topologies of the graphene layer are so different with total energies almost equal (difference of 0.007\%).   

\begin{figure}[htb]
\begin{minipage}{0.45\linewidth}
\includegraphics[width=0.95\linewidth]{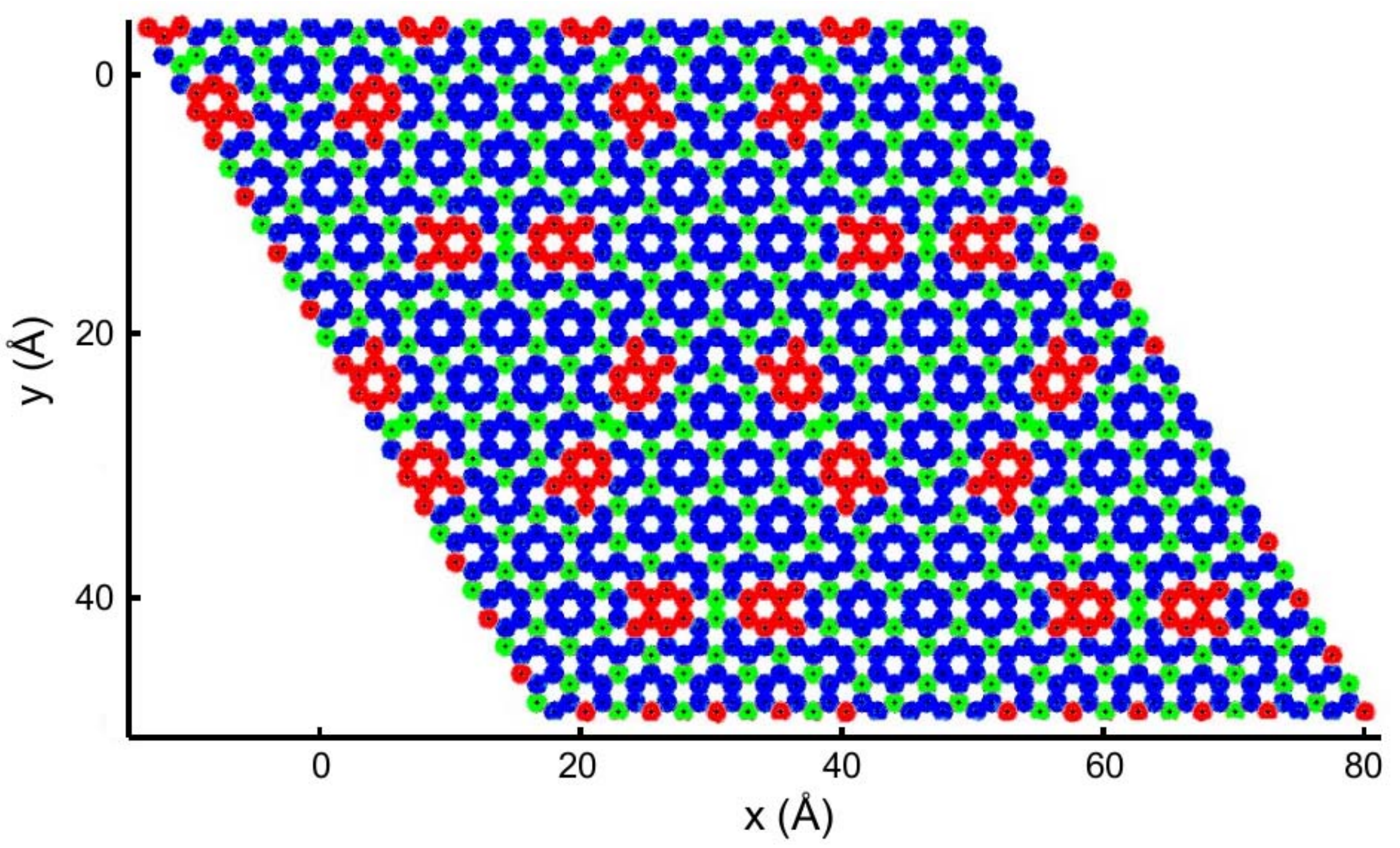}
\caption{\label{FigIntuition255} Strongly sticked three-level structure of the graphene layer. Blue dots are for C atoms closed to the surface, red dots for those far from the surface and green ones for intermediate values of $z$. Mean gap equal to 2.55 \AA.}
\end{minipage}
\begin{minipage}{0.45\linewidth}
\includegraphics[width=0.95\linewidth]{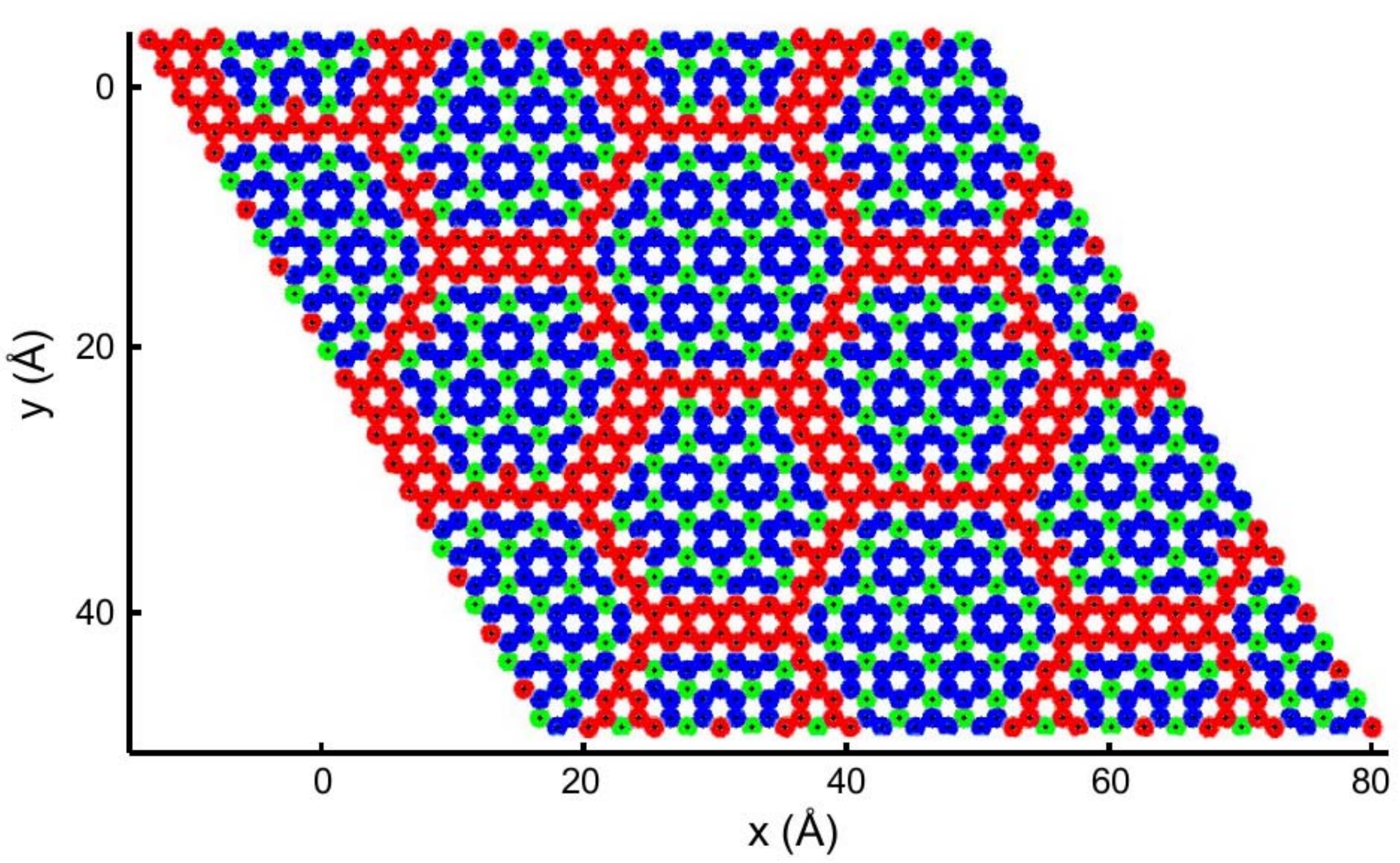}
\caption{\label{FigIntuition258} Moderately sticked three-level structure of the graphene layer. Mean gap equal to 2.58 \AA. Same color code than in Fig. \ref{FigIntuition255}.}
\end{minipage}
\begin{minipage}{0.45\linewidth}
\includegraphics[width=0.95\linewidth]{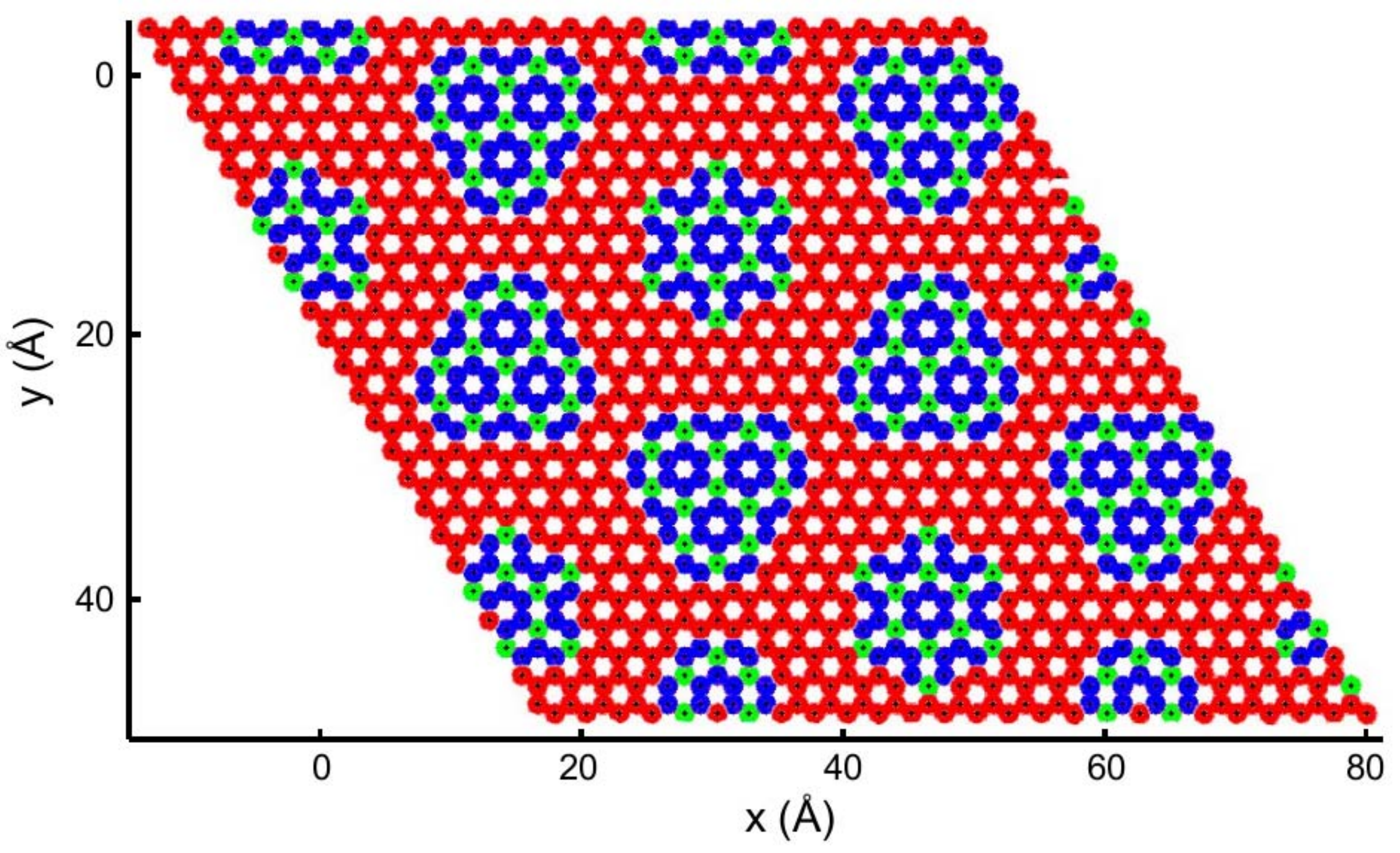}
\caption{\label{FigIntuition262}  Weakly sticked three-level structure of the graphene layer. Mean gap equal to 2.62 \AA. Same color code than in Fig. \ref{FigIntuition255}.}
\end{minipage}
\end{figure}

\begin{figure}[htb]
\begin{minipage}{0.45\linewidth}
\includegraphics[width=0.95\linewidth]{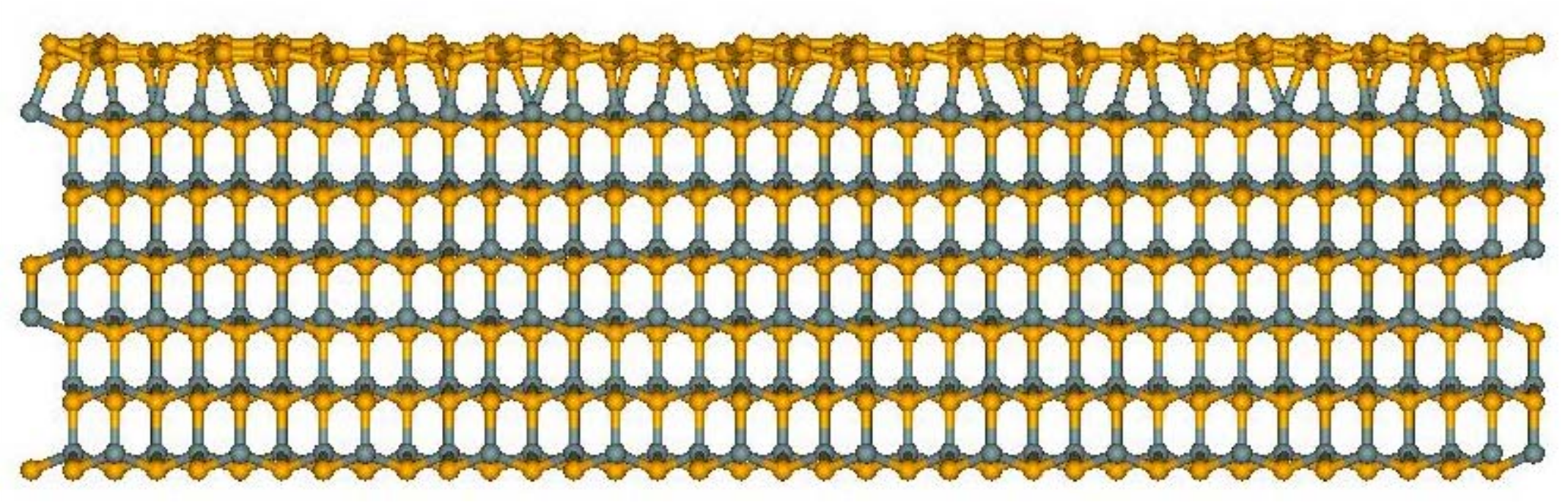}
\caption{\label{FigSolNRel} Cross view of the system presented in Fig. \ref{FigIntuition255}.}
\end{minipage}
\begin{minipage}{0.45\linewidth}
\includegraphics[width=0.95\linewidth]{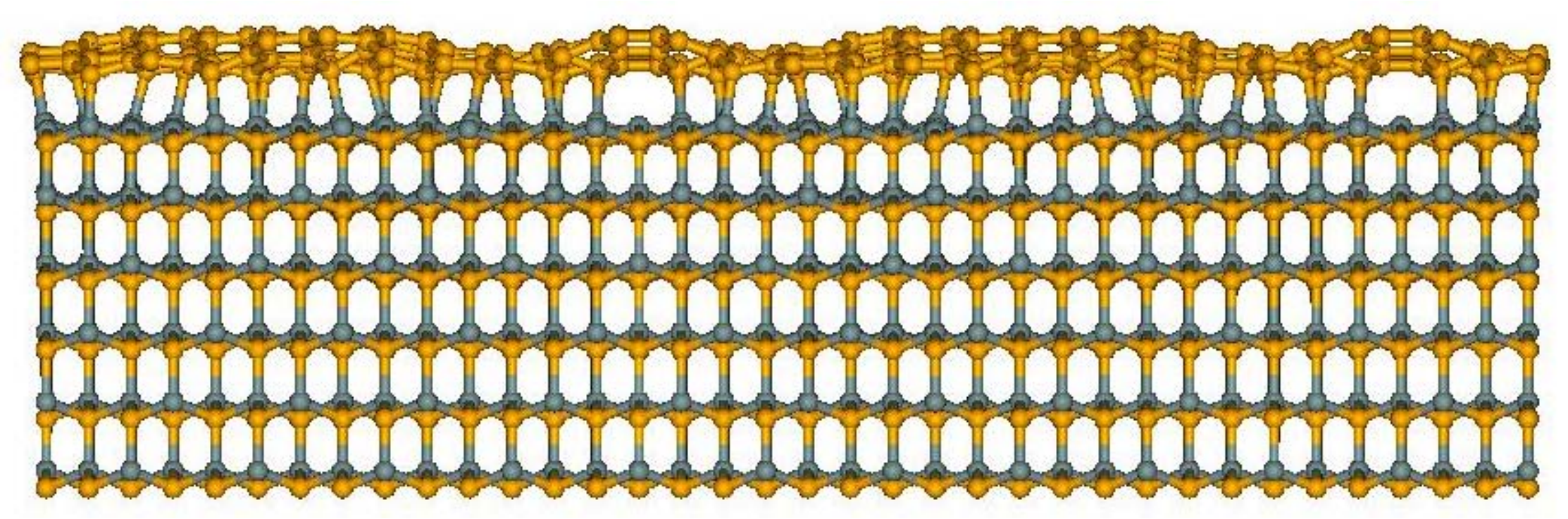}
\caption{\label{FigSolRel} Cross view after relaxation of the system presented in Fig. \ref{FigIntuition255}.}
\end{minipage}
\end{figure}

Therefore we decided to refine the construction of the initial structure to facilitate the relaxation. Indeed, it seems obvious that the multiple possible binding of some of the C atoms of the graphene layer to the Si of the surface is too difficult to handle by the procedure of conjuguate gradients. Since the choice of the initial configuration is of primary importance, 3 initial structures have been specifically built as illustrated in Figs. \ref{FigIntuition255}  to \ref{FigIntuition262}
In these structures, the graphene layer is no longer planar, but the C atoms are set at 3 different heights $z$. The C atoms initially close to a Si at the surface are set at the lower $z$ to strengthen the bonds, their immediate surrounding to an intermediate $z$ and the remaining atom to a higher $z$. By varying the initial mean gap between graphene and the surface, the layer is more or less blue, i.e. more or less sticked to the SiC surface. Fig. \ref{FigIntuition255} corresponds to a mean gap of 2.55 \AA, Fig. \ref{FigIntuition258} to 2.58 \AA\ and Fig. \ref{FigIntuition262} to 2.62 \AA. A cross view of the system shown on Fig. \ref{FigIntuition255} is also given in Fig. \ref{FigSolNRel} and evidences the slight separation along $z$ of the three levels.

\begin{figure}[htb]
\includegraphics[width=0.45\linewidth]{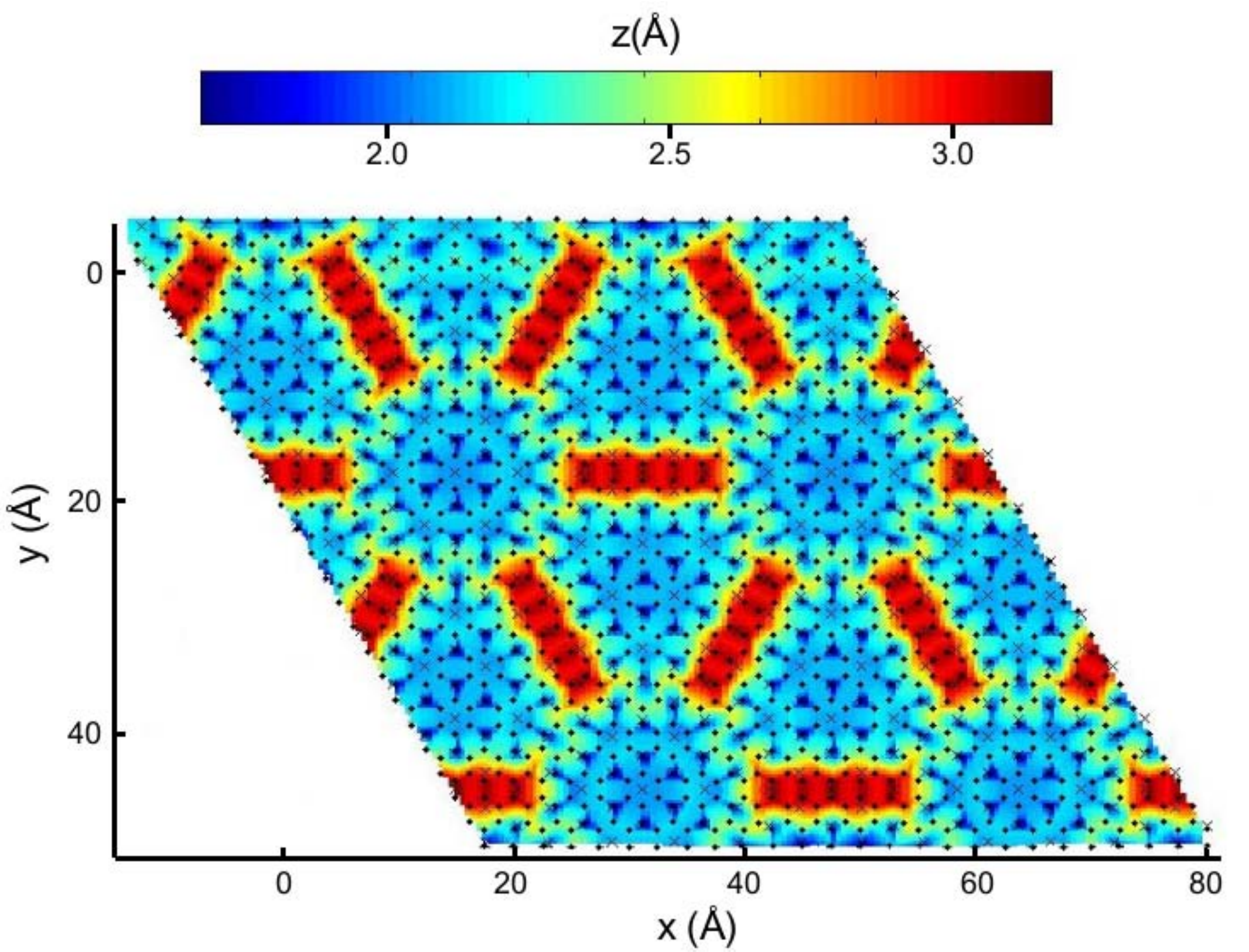}
\caption{\label{FigBest} Color map of the positions along $z$ of the C atoms (black dots) in graphene. Most stable configuration. Blue (red) means near (far from) the SiC surface. Light crosses indicate the position of the Si atoms of the surface.}
\end{figure}

After relaxation, the graphene sheet presents a pronounced buckling of magnitude typically equal to 1.5 \AA\ (Fig. \ref{FigSolRel}).
The energy obtained with the three 3-levels systems is smaller than the minimum obtained using a planar sheet as a starting point (Fig. \ref{FigCherche}). The most stable structure is obtained using the highly-sticked system presented in Fig. \ref{FigIntuition255}. A color map of the positions along $z$ of the C atoms in graphene is given in Fig. \ref{FigBest} for the corresponding system.
Low regions form circular and triangular patterns separated by rods at higher $z$. These rods are necessary to release the misfit of 3.0 \% between the graphene sheet and the substrate. The duplication of the elementary pattern results in the formation of an hexagonal structure formed of a central disk surrounded by six triangles separated by rods distributed in a six-branch star. The cross view shown in Fig. \ref{FigSolRel} suggests that the C atoms in these rods are not bonded to the surface, an assumption that will be discussed in the next section. By comparing the initial structure (Fig. \ref{FigIntuition255}) to the relaxed one, we guess that the half rods already present initially extend to form complete rods during the relaxation.  
\section{Discussion}
The result presented in Fig. \ref{FigBest} has several interesting features. First, it has the lowest energy ever found in our various calculations. Moreover, the structure presents a high level of order, in particular compared to solutions such as the one presented in Fig. \ref{Fig1.63}. The symetry level is also the highest, and the $(6\sqrt{3}\times6\sqrt{3})R30^{\circ}$ and $(6\times6)$ periodicities already observed by scanning tunneling microscopy \cite{Riedl07} are visible in Fig. \ref{FigBest}. This solution presents all the features expected for the equilibrium state.

The gap between the SiC surface and the graphene layer, defined in this case as the distance between the upper Si atom of the surface and the lower C of the graphene, is equal to 1.68 \AA. This gap is smaller than the equilibrium distance of the Si-C bond with EDIP, i. e. 1.91 \AA, which is due to the formation of leaning connections between atoms that are not at the vertical of each other. The higher C atoms in graphene are at $z=3.17$ \AA, a value to compare to the cutoff radius of the Si-C interactions with EDIP equal to 2.67 \AA. As foreseen in the previous section, the C atoms in the rods are indeed not bonded to the surface, their location is only fixed by $sp2$ interactions with their neighbours in the buffer layer. The amplitude of graphene buckling is equal to 1.50 \AA. The energy per C atom in graphene is equal to -7.42 eV. This value is lower than the cohesive energy in graphite (-7.38 eV), which is due to the formations of Si-C bonds. Moreover, the impact of the substrate thickness, which could be important due to the stress of the bonded graphene in misfit has been studied. The test has been performed on the energy per graphene atom since the total number of atom is not constant. It has been obtained a variation of only 0.0056\% from 4 to 6 SiC bilayers, and 0.0005 \% from 6 to 15 bilayers. Obviously 4 SiC bilayers are enough to absorb the strain.

\begin{figure}[htb]
\includegraphics[width=0.45\linewidth]{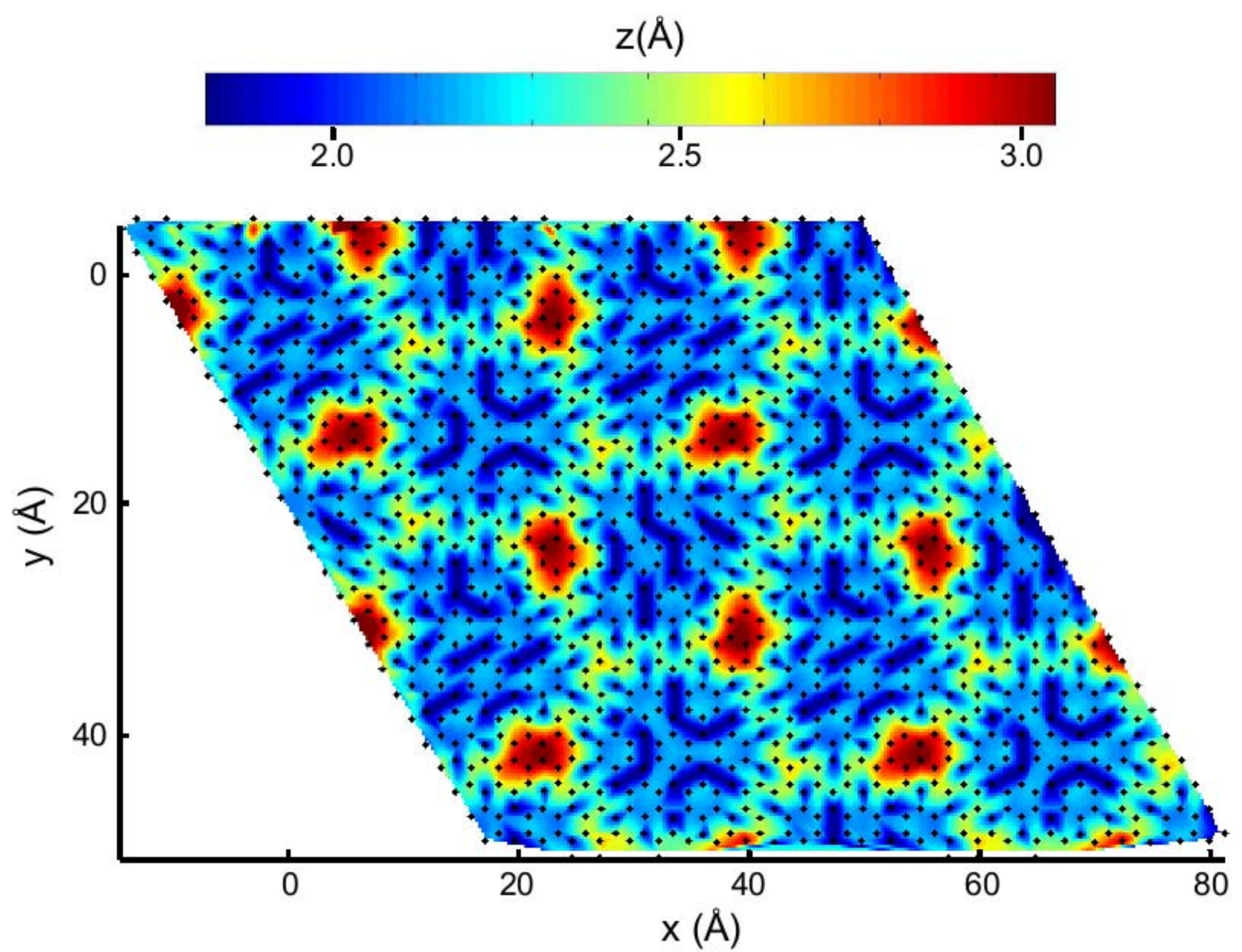}
\caption{\label{FigLaurence} Color map of the positions along $z$ of the C atoms (black dots) in graphene. Result of the \emph{ab initio} calculation presented in Ref. \onlinecite{Varchon08}. Blue (red) means near (far from) the SiC surface.}
\end{figure}

The present result obtained with the EDIP empirical potential is afterwards compared to the result of an \emph{ab initio} calculation \cite{Varchon08} (Table \ref{TabComp}).
\begin{table}
\begin{tabular}{cccc}
\hline\hline
&\emph{ab initio} & Empirical potential\\
\hline
Interatomic interations &GGA\cite{Perdew86} (code VASP\cite{Kresse94}) &Modified version of EDIP\cite{Lucas06}\\
\hline
Misfit graphene/SiC &0.7\% &3.0 \%\\
\hline
Basic cell& 1310 atoms, 4 SiC bilayers,& 6536 atoms, 6 SiC bilayers,\\
characteristics & H-passivation at bottom&  frozen atoms at bottom\\
\hline
Convergence& Residual forces & Variation in energy\\
criterium&  $<$ 15 meV/\AA &  $<$ tenths of $\mu$eV/atom\\
\hline
Number of steps &&Tenths to\\
for relaxation && hundreds\\
\hline\hline
\end{tabular}
\caption{\label{TabComp} Main features of the \emph{ab initio} and empirical potential calculations.}
\end{table}
 The structure is duplicated and a color map is applied to visualize the variations in height of the C in graphene as in Fig. \ref{FigBest}. The result is presented in Fig. \ref{FigLaurence}.
The structure is less ordered, and an elementary pattern does not clearly stand out, although the hexagonal structure is undoubtly appearent at a larger scale. The rods are smaller and the low regions are not so regular but circular forms and quasi triangular ones can be distinguished. The difficulty to find the minimum is probably here again at the origin of the irregularities of the structure, but the pattern globally corresponds to the result of the EDIP calculation. 
\begin{figure*}[htb]
\centering
\includegraphics[width=8cm]{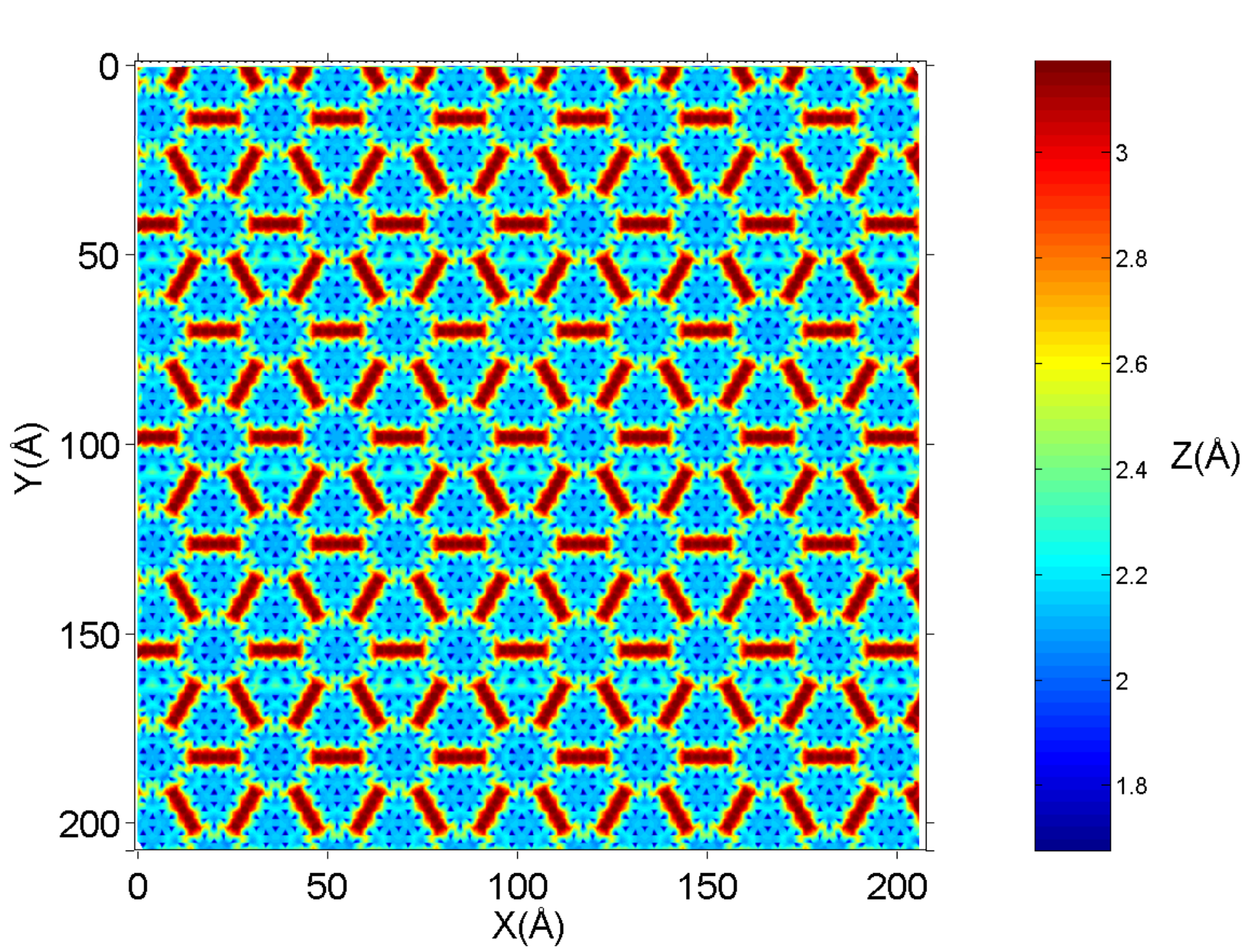}
\includegraphics[width=8cm]{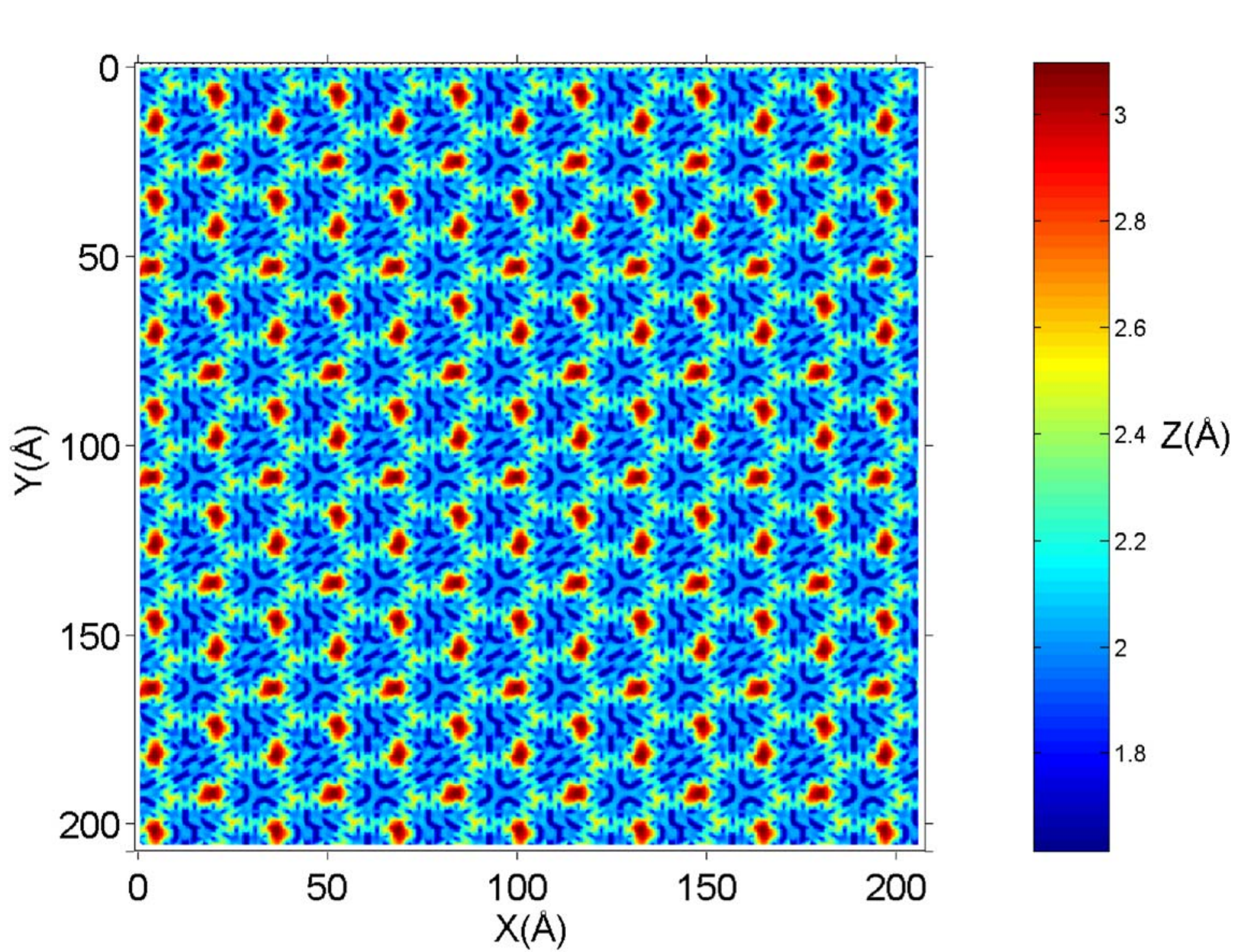}
\includegraphics[width=7cm]{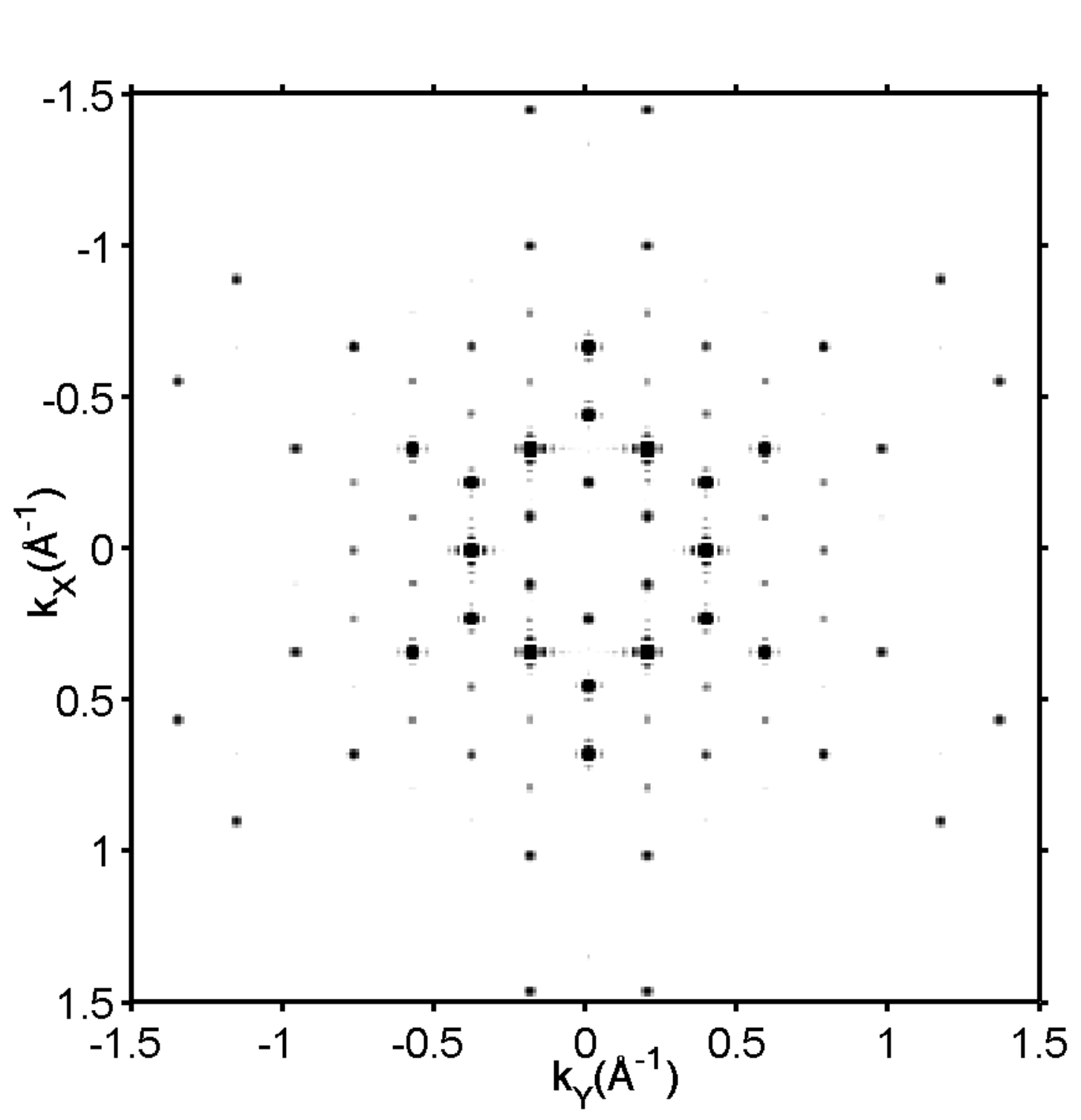}
\hspace{1cm}
\includegraphics[width=7cm]{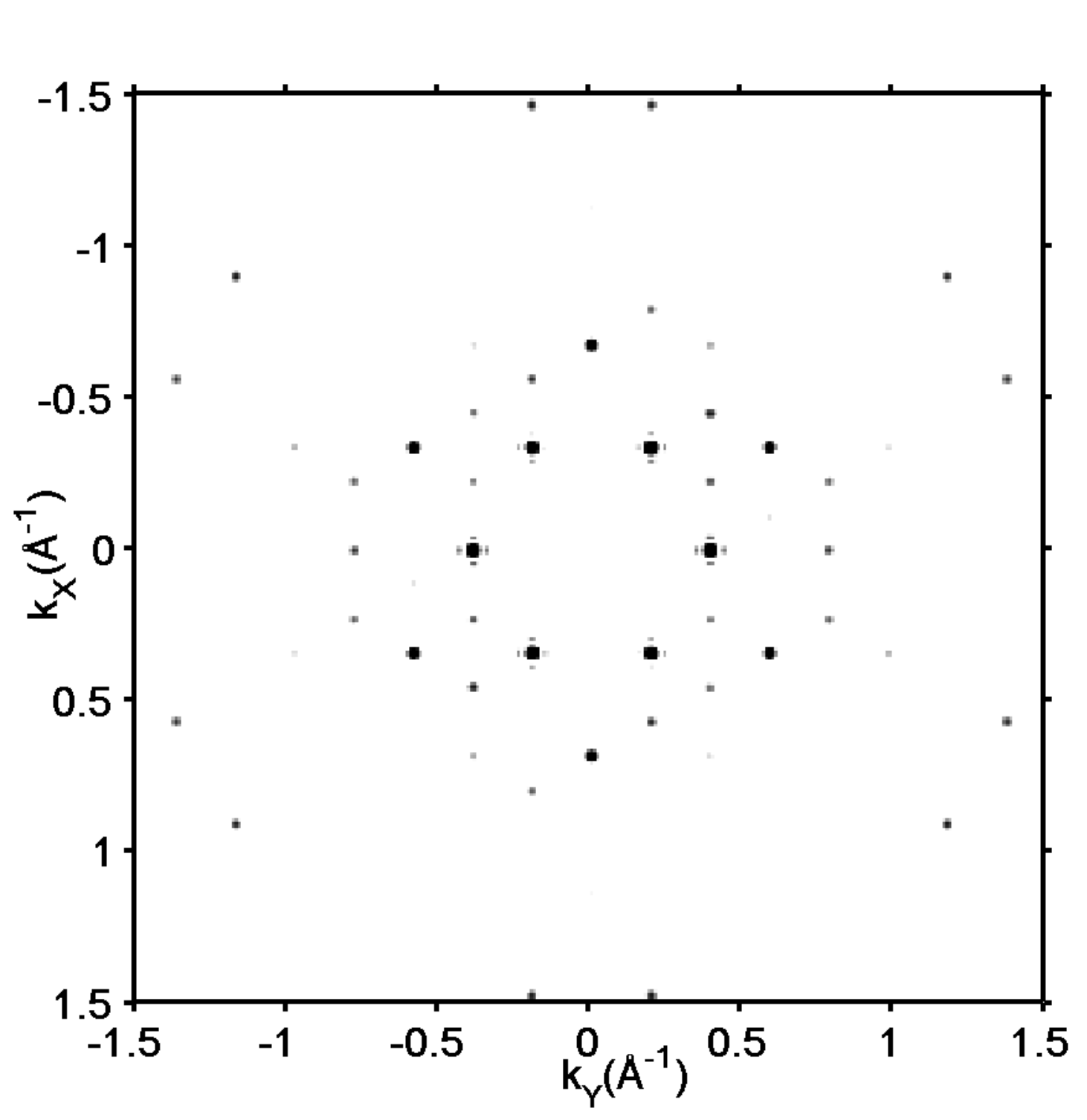}
\caption{\label{FigLargeScale} Top: color map on the poisitons of C in graphene. Bottom: Fourier transform of the top color map, zoom at low $k$ to emphasize the large scale order. Left: EDIP. Right: \emph{ab initio}\cite{Varchon08}.}
\end{figure*}
The atomic positions along $z$ and the corresponding Fourier transforms are drawn at larger scale in Fig. \ref{FigLargeScale}. The Fourier transforms evidence some particular features in common (hexagon at 0.3 \AA$^{-1}$ for the 6$\times$6-SiC modulation and 0.7 \AA$^{-1}$ for a local symmetry of the more or less circular grains at low $z$) and the higher level of symmetry obtained with the present minimisation (second hexagon at 0.3 \AA$^{-1}$ coming from the 30$^{\circ}$-rotated 6$\times$6-SiC periodicity and central hexagon at 0.2 \AA$^{-1}$ due to the $(6\sqrt{3}\times6\sqrt{3})R30^{\circ}$ elementary pattern, at least more intense in the present calculation). 

The buckling amplitude of the graphene layer is equal to 1.2 \AA\ to be compared to 1.5 \AA\ for EDIP, for respectively 0.7 and 3.0 \% of graphene-SiC misfit. The space between graphene and the surface is equal to 1.8 \AA, versus 1.7 \AA\ for EDIP. 

The global agreement of the two calculations leads us to the conclusion that this empirical potential is perfectly suited for the study of graphene on SiC. The level of order of the superstructure is even more pronounced, due to the methodology developed to find the minimum despite an empirical physical description of atomic interactions. If in this case, \emph{ab initio} calculations were still affordable, the use of an empirical description will be essential to study larger systems with surface steps or perform molecular dynamics simulations at a given temperature in order to investigate the mechanism of Si sublimation that results in the formation of graphene. 

\section{Conclusion} 
In conclusion, the bonding of the graphene buffer layer to the Si-terminated surface of SiC has been studied using a modified version of the interatomic potential EDIP. The difficulty to find the equilibrium configuration has been emphasized, and a dedicated procedure has been developed in order to produce a starting configuration on the way to the minimum. The result presents high level of order and symmetry, with an hexagonal pattern formed of sticked circular and triangular regions and unbonded rods to release the misfit. The structure is coherent with the result obtained using an \emph{ab initio} description and presents all the symetries identified in experiments. The physical description is therefore validated for further study on the graphitization process.

\end{document}